\documentclass[9pt,conference]{IEEEtran}
\IEEEoverridecommandlockouts

\usepackage{cite}
\usepackage{amsmath,amssymb,amsfonts}
\usepackage{algorithmic}
\usepackage{graphicx}
\usepackage{textcomp}
\usepackage{xcolor}
\usepackage{lipsum}
\usepackage{CJKutf8}
\usepackage{booktabs}
\usepackage{arydshln}
\usepackage{multirow}
\usepackage{utfsym}
\usepackage{url}

\def\BibTeX{{\rm B\kern-.05em{\sc i\kern-.025em b}\kern-.08em
    T\kern-.1667em\lower.7ex\hbox{E}\kern-.125emX}}

\begin{document}

\title{SLAM-AAC: Enhancing Audio Captioning with Paraphrasing Augmentation and CLAP-Refine through LLMs
\\
\thanks{Co-first author$^\ast$. Corresponding author$^\dag$.}
}

\author{
\IEEEauthorblockN{Wenxi Chen$^\ast$}
\IEEEauthorblockA{\textit{X-LANCE Lab} \\
\textit{Shanghai Jiao Tong University}\\
Shanghai, China \\
1029713857@sjtu.edu.cn}
\and
\IEEEauthorblockN{Ziyang Ma$^\ast$}
\IEEEauthorblockA{\textit{X-LANCE Lab} \\
\textit{Shanghai Jiao Tong University}\\
Shanghai, China \\
zym.22@sjtu.edu.cn}
\and
\IEEEauthorblockN{Xiquan Li}
\IEEEauthorblockA{\textit{X-LANCE Lab} \\
\textit{Shanghai Jiao Tong University}\\
Shanghai, China \\
mtxiaoxi55@sjtu.edu.cn}
\and
\IEEEauthorblockN{Xuenan Xu}
\IEEEauthorblockA{\textit{X-LANCE Lab} \\
\textit{Shanghai Jiao Tong University}\\
Shanghai, China \\
wsntxxn@gmail.com}
\and
\IEEEauthorblockN{Yuzhe Liang}
\IEEEauthorblockA{\textit{X-LANCE Lab} \\
\textit{Shanghai Jiao Tong University}\\
Shanghai, China \\
l.yzzzz@sjtu.edu.cn}
\and
\IEEEauthorblockN{Zhisheng Zheng}
\IEEEauthorblockA{\textit{X-LANCE Lab} \\
\textit{Shanghai Jiao Tong University}\\
Shanghai, China \\
zzs666@sjtu.edu.cn}
\and
\IEEEauthorblockN{Kai Yu}
\IEEEauthorblockA{\textit{X-LANCE Lab} \\
\textit{Shanghai Jiao Tong University}\\
Shanghai, China \\
kai.yu@sjtu.edu.cn}
\and
\IEEEauthorblockN{Xie Chen$^\dag$}
\IEEEauthorblockA{\textit{X-LANCE Lab} \\
\textit{Shanghai Jiao Tong University}\\
Shanghai, China \\
chenxie95@sjtu.edu.cn}
}

\maketitle

\begin{abstract}
Automated Audio Captioning (AAC) aims to generate natural textual descriptions for input audio signals.
Recent progress in audio pre-trained models and large language models (LLMs) has significantly enhanced audio understanding and textual reasoning capabilities, making improvements in AAC possible.
In this paper, we propose SLAM-AAC to further enhance AAC with paraphrasing augmentation and CLAP-Refine through LLMs. 
Our approach uses the self-supervised EAT model to extract fine-grained audio representations, which are then aligned with textual embeddings via lightweight linear layers.
The caption generation LLM is efficiently fine-tuned using the LoRA adapter.
Drawing inspiration from the back-translation method in machine translation, we implement paraphrasing augmentation to expand the Clotho dataset during pre-training. 
This strategy helps alleviate the limitation of scarce audio-text pairs and generates more diverse captions from a small set of audio clips.
During inference, we introduce the plug-and-play CLAP-Refine strategy to fully exploit multiple decoding outputs, akin to the n-best rescoring strategy in speech recognition.
Using the CLAP model for audio-text similarity calculation, we could select the textual descriptions generated by multiple searching beams that best match the input audio. 
Experimental results show that SLAM-AAC achieves state-of-the-art performance on Clotho V2 and AudioCaps, surpassing previous mainstream models.
\end{abstract}


\section{Introduction}
\label{sec:intro}
Automated audio captioning (AAC) is a challenging multimodal task aimed at generating natural textual descriptions from audio data. 
Unlike conventional audio understanding tasks such as audio tagging (AT), AAC requires systems to not only comprehend the content of audio clips but also to align textual and acoustic modalities, ultimately producing coherent and linguistically fluent descriptions \cite{xu2023beyond}.

In AAC tasks, sequence-to-sequence (seq2seq) architectures are widely adopted, where audio encoders extract acoustic representations, and language models utilize these representations to generate captions auto-regressively. 
Traditional methods \cite{cho2023hyu} often rely on supervised models, such as PANNs \cite{kong2020panns}, for audio feature extraction. 
Recently, self-supervised pre-trained models like BEATs \cite{chen2022beats} have been integrated into AAC systems \cite{wu2023beats,tang2024extending}, resulting in notable performance improvements.
In this work, our proposed SLAM-AAC\footnote{SLAM-AAC is a subproject of SLAM-LLM~\cite{ma2024embarrassingly}, where SLAM stands for \textbf{S}peech, \textbf{L}anguage, \textbf{A}udio and \textbf{M}usic. The project is open-sourced and available at \scriptsize{\url{https://github.com/X-LANCE/SLAM-LLM}.}} model employs the Efficient Audio Transformer (EAT) \cite{chen2024eat}, a self-supervised pre-trained model that achieves state-of-the-art performance in audio tagging tasks \cite{gemmeke2017audio}, as the audio encoder to extract more fine-grained audio representations.
To enhance computational efficiency and improve alignment between audio and text embeddings, we use lightweight linear layers to downsample the 50Hz audio representations to approximately 10Hz.
For text decoding, the advent of large language models (LLMs)  \cite{chung2024scaling,touvron2023llama,radford2019language,vicuna2023} has demonstrated superior understanding and reasoning capabilities compared to smaller models  \cite{lewis2019bart}, leading to more fluent and natural text generation for AAC tasks. 
Consequently, we adopt the large language model Vicuna  \cite{vicuna2023} as the text decoder, which attends to the aligned audio and text representations to generate corresponding textual descriptions. 
To further enhance training efficiency, SLAM-AAC integrates LoRA \cite{hu2021lora} adapters for parameter-efficient fine-tuning (PEFT) of the LLM, while keeping EAT frozen and training only the alignment layers.

In AAC tasks, the scarcity of high-quality audio-text paired datasets presents a significant challenge, highlighting the importance of effective data augmentation techniques to improve model performance.
In SLAM-AAC, we employ both audio and text augmentation during model training.
For audio augmentation, we apply SpecAugment \cite{park2019specaugment} to proportionally mask the audio mel-spectrogram in both time and frequency dimensions, enhancing the model's robustness.
In previous AAC methods \cite{cho2023hyu,koizumi2020ntt}, text augmentation has involved the use of WordNet \cite{miller1995wordnet} for synonym replacement or altering words with low TF-IDF scores to preserve the informativeness of key terms.
Instead of word-level substitution, we introduce sentence-level augmentation by generating paraphrases through back-translation \cite{sennrich2015improving} for each audio caption.
This approach expands the Clotho \cite{drossos2020clotho} training set, which has limited data compared to other AAC datasets like AudioCaps \cite{kim2019audiocaps} and WavCaps \cite{mei2023wavcaps}, thereby increasing the diversity and complexity of the pre-training data for SLAM-AAC.

In automatic speech recognition (ASR), the n-best rescoring strategy \cite{mikolov2010recurrent,chen2015recurrent} is widely used to reduce word error rates (WER), where a language model is trained to score and select the most accurate result from a list of n-best decoded candidates.
Building on this concept, we introduce the plug-and-play CLAP-Refine strategy to enhance text decoding.
The Contrastive Language-Audio Pre-training model (CLAP) \cite{wu2023large} projects text and audio features into a shared space using contrastive pre-training, enabling the calculation of text-audio similarity.
By evaluating the similarity of candidate captions generated through multiple beam search decoding with the input audio, the proposed CLAP-Refine selects the highest-scoring caption as output.
This method enables SLAM-AAC to effectively leverage different beam search results, providing the best matching text description for the input audio.

\begin{figure}[t]
\begin{minipage}[b]{1.0\linewidth}
  \centering
  \centerline{\includegraphics[width=8.5cm]{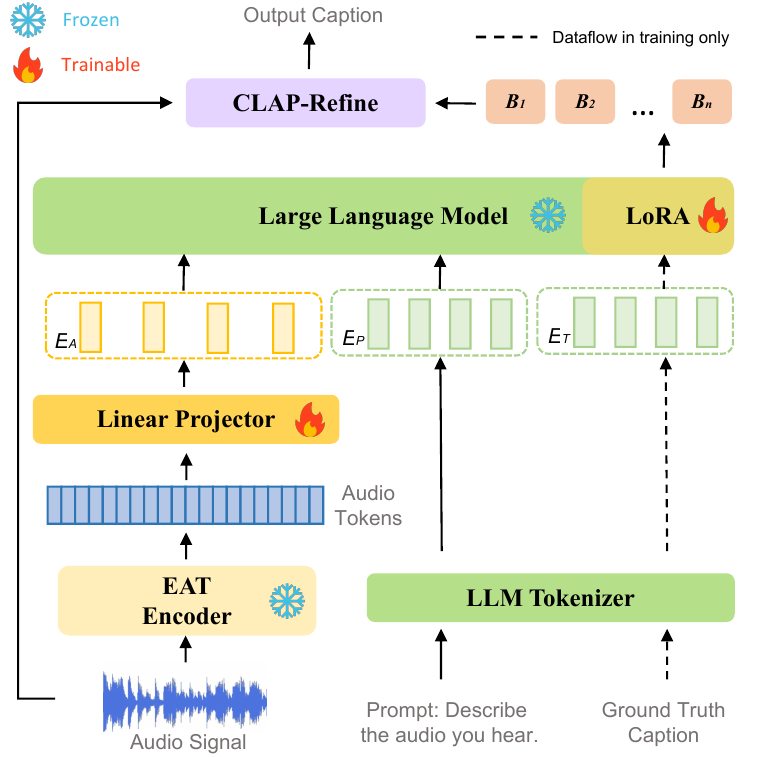}}
   \caption{Overview of the SLAM-AAC system. We use the frozen EAT to extract fine-grained audio representations, which are then downsampled and aligned with text embeddings via a linear projector.  The LLM for decoding generates text based on these concatenated representations and is efficiently fine-tuned using LoRA. During inference, multiple candidate captions are generated through various beam searches, with the most audio-aligned textual description selected as the final output using the CLAP-Refine strategy. Here, $B_n$ denotes the candidate generated using a beam size of $n$ in decoding.}
  \label{fig:SLAM-AAC model}
\end{minipage}
\end{figure}

SLAM-AAC achieves state-of-the-art performance across all AAC metrics on both the AudioCaps and Clotho evaluation split. We will open-source our model weights and code.
We open-source our model weights, code, and the augmented Clotho dataset.\footnote{\scriptsize{\url{https://github.com/X-LANCE/SLAM-LLM/blob/main/examples/slam_aac}}}

\section{SLAM-AAC}

\subsection{Network Architecture}
Following existing AAC models \cite{wu2023beats,tang2024extending,chen2024sjtu}, SLAM-AAC employs a sequence-to-sequence framework, as illustrated in Fig. \ref{fig:SLAM-AAC model}.

For audio encoding, SLAM-AAC employs the Transformer-based model EAT \cite{chen2024eat} as its acoustic feature extractor.
EAT is an efficient self-supervised pre-trained model that utilizes masked language modeling \cite{devlin2018bert} as its pretext task within a self-distilled framework. 
It demonstrates significant performance improvement over both supervised models like PANNs \cite{kong2020panns} and self-supervised models like BEATs \cite{chen2022beats}, particularly in audio classification tasks such as AS-2M and AS-20K \cite{gemmeke2017audio}. 
In our experiments, we utilize the EAT-base\footnote{EAT-base\_epoch30 (fine-tuned on AS-2M), with 88M parameters} model, which has been pre-trained and fine-tuned on the AudioSet dataset, to extract audio representations.
EAT resamples the input waveforms to a 16kHz sample rate and transforms them into 128-dimensional Mel-frequency bands using a 25ms Hanning window with a 10ms shift. 
These mel-spectrograms are then converted into 2D patch embeddings through a CNN encoder, followed by feature extraction via a 12-layer ViT-B model \cite{dosovitskiy2020image}.
The resulting audio representations, $E_a$, have a frequency of approximately 50Hz. 
To facilitate the alignment of audio-text modalities and reduce the feature sequence length, we apply lightweight 2-layer linear projections for a 5x downsampling, converting audio tokens $E_a$ into $E_A$ at approximately 10Hz.

To enhance text generation, our approach employs the LLM Vicuna\footnote{\scriptsize{\url{https://huggingface.co/lmsys/vicuna-7b-v1.5}}}\cite{vicuna2023} as the decoder. 
As shown in Fig. \ref{fig:SLAM-AAC model}, SLAM-AAC processes textual prompts (e.g., ``Describe the audio you hear") and ground truth captions using Vicuna's tokenizer with a 32K vocabulary, generating the corresponding text embeddings \( E_P \) and \( E_T \).
During training, the ground truth caption is incorporated using teacher-forcing, and the joint embeddings \( E_J \) attended by the decoder are formed by concatenating these embeddings as follows:

\[
E_J = 
\begin{cases} 
[E_A; E_P; E_T] & \text{during training} \tag{1}\\
[E_A; E_P] & \text{during inference} 
\end{cases}
\]

Let \( |E_T| \) denote the length of the ground truth text embeddings, and let \( E_{T,t} \) represent the \( t \)-th token of \( E_T \). 
The training objective of our system is the cross-entropy loss, defined as:

\[
\mathcal{L}_{CE} = -\frac{1}{|E_T|} \sum_{t=1}^{|E_T|} \log p(E_{T,t} | E_{T,1:t-1}, E_J) \tag{2}
\]

To improve training efficiency, we employ LoRA \cite{hu2021lora} to fine-tune the LLM. 
As a result, only the linear projections and the LoRA adapter are trainable, while the rest of the model remains frozen.

\subsection{Paraphrasing Augmentation in AAC}
\label{subsec:paraphrasing}
In AAC datasets, audio annotations are typically produced by a limited number of annotators, leading to consistent annotation styles and a restricted vocabulary, which can hinder the overall diversity of the dataset. 
Specifically, the Clotho dataset \cite{drossos2020clotho}, part of the SLAM-AAC training data, features a variety of audio clips of varying lengths. 
However, compared to other datasets such as AudioCaps \cite{kim2019audiocaps} and WavCaps \cite{mei2023wavcaps}, the Clotho training set is relatively small, with only 3,839 audio clips, each paired with five annotated captions. 
To mitigate this limitation, we innovatively introduced the paraphrasing augmentation based on back-translation \cite{sennrich2015improving} to expand the audio-text pairs in Clotho. This method aims to enrich the dataset with diverse captions and enhance the model's generalizability.

\begin{table}[htbp]
    \centering
    \caption{An example of back-translation paraphrasing in Clotho.}
    {
    \begin{tabular}{|p{2.8cm}|p{4.8cm}|}
        \hline
        \textbf{Original Annotation} \par \textbf{(English)} & A person is very carefully wrapping a gift for someone else. \\ \hline
        \textbf{Translation} \par \textbf{(Chinese)} & \begin{CJK}{UTF8}{gbsn}\begin{fontsize}{8pt}{0pt}一个人正在非常小心地为别人包装礼物。\end{fontsize}\end{CJK} \\ \hline
        \textbf{Back-Translation} \par \textbf{(English)}   & Someone is wrapping a present for someone else with great care. \\ \hline
    \end{tabular}
    }
    \label{tab:backtranslation}
\end{table}

Table \ref{tab:backtranslation} provides an example of back-translation applied to Clotho. 
In our experiments, we used the Google Translate API\footnote{\scriptsize{\url{https://cloud.google.com/translate}}} to translate each English caption into Chinese and then back into English, thereby generating five additional annotations per audio clip. 
Consequently, the total vocabulary of the annotations expanded from 7,454 to 10,453 words. 
This paraphrasing technique enables more extensive sentence restructuring compared to previous word-level text augmentation \cite{cho2023hyu,koizumi2020ntt} while preserving the original semantics.

\begin{table*}[htbp]
\centering
\caption{
Performance comparison of AAC models on Clotho and AudioCaps evaluation split. 
}
\resizebox{1\linewidth}{!}{
\begin{tabular}{cccccccccccccccc}
\toprule
\multirow{2.5}{*}{\textbf{Model}} & \multirow{2.5}{*}{\textbf{Pre-training Data}}  & \multicolumn{6}{c}{\textbf{Clotho Evaluation} (\%)} & & \multicolumn{6}{c}{\textbf{AudioCaps Evaluation} (\%)} \\
\cmidrule(lr){3-8} \cmidrule(lr){10-15}
 &  &  MT & CD & SC & SD & SF & FS && MT & CD & SC & SD & SF & FS  \\
\midrule
EnCLAP-large \cite{kim2024enclap} & AC+CL & 18.6 & 46.4 & 13.3 & 29.9 & 28.9$^{\mathrm{a}}$  & 50.7$^{\mathrm{a}}$ && 25.5 & 80.3 & 18.8 & 49.5 & 49.9$^{\mathrm{a}}$ & 65.5$^{\mathrm{a}}$
\\
 WavCaps$^{\mathrm{b}}$ \cite{mei2023wavcaps} & AC+CL+WC & 18.5 & 48.8 & 13.3 & 31.0 &  29.6$^{\mathrm{a}}$ & 50.1$^{\mathrm{a}}$ && 25.0 & 78.7 & 18.2 & 48.5 & 48.3$^{\mathrm{a}}$ & 64.2$^{\mathrm{a}}$
 \\
Wu et al. \cite{wu2023beats} & AC+CL$_{C}$ & 19.3 &  50.6  &  14.6  & 32.6 & 32.6 &  53.6  &&  -  & -  &  -  & - & - & -  
\\
Tang et al. \cite{tang2024extending} & AC+CL+WC+LS+GS & - &  -  &  -  & 31.8 & - &  -  &&  -  & -  &  -  & 50.6 & - & -  
\\
\hdashline
SLAM-AAC (ours) & AC+CL$_{P}$+WC+MA & \textbf{19.7} & \textbf{51.5} & \textbf{14.8} & \textbf{33.2} &  \textbf{33.0} & \textbf{54.0} && \textbf{26.8} & \textbf{84.1} & \textbf{19.4} & \textbf{51.8} & \textbf{51.5} &\textbf{66.8}   \\
\bottomrule
\multicolumn{16}{l}{\textbf{Pre-training datasets:} AudioCaps (AC), Clotho (CL), WavCaps (WC), MACS (MA), LibriSpeech (LS), and GigaSpeech (GS).} \\
\multicolumn{16}{l}{\textbf{Metrics:} METEOR (MT), CIDEr (CD), SPICE (SC), SPIDEr (SD), SPIDEr-FL (SF), and FENSE (FS).} \\
\multicolumn{16}{l}{CL$_{C}$ and CL$_{P}$ denote the Clotho training set augmented with the ChatGPT Mix-up method and our paraphrasing approach, respectively.} \\
\multicolumn{16}{l}{$^{\mathrm{a}}$ For open-source models, we evaluated metrics not reported in the original papers using our evaluation split.} \\
\multicolumn{16}{l}{$^{\mathrm{b}}$ The best performance of WavCaps on both datasets is selected for comparison (model architectures may differ).} 
\end{tabular}
}
\label{tab:main}
\end{table*}

\subsection{CLAP-Refine for Text Decoding}
Traditional AAC systems often employ beam search \cite{cho2023hyu} or nucleus sampling \cite{wu2023beats} for text decoding. 
However, beam search focuses primarily on maximizing the text decoder's score during decoding, often neglecting the alignment between the generated text and the audio embeddings. 
On the other hand, nucleus sampling can lead to unstable outputs, usually necessitating multiple decoding attempts and extensive post-processing to reach optimal results \cite{wu2023beats}.

To address these limitations, we introduce an innovative, plug-and-play strategy named CLAP-Refine, which enhances decoding results by effectively leveraging multiple beam searches as a post-processing step. 
CLAP \cite{wu2023large} constructs an implicit audio-text multi-modal semantic space through contrastive learning. 
It employs dual encoders to independently process text and audio data, generating corresponding representations \(C_A\) for audio and \(C_T\) for text within a shared representation space. 
The model could measure the alignment of text-audio pairs using cosine similarity, defined as:

\[
\text{Similarity}(C_A, C_T) = \frac{C_A \cdot C_T}{\|C_A\| \|C_T\|}
\tag{3}
\]

For inference, SLAM-AAC first generates the most probable sentences with different beam sizes for the same input audio, resulting in a set of candidate captions $B_1, B_2, ..., B_n$, as illustrated in Fig. \ref{fig:SLAM-AAC model}. 
The CLAP-Refine strategy then employs the CLAP model to compute similarity scores between each candidate caption and the input audio. 
These scores are used to rerank the candidates, refining the outputs to prioritize those most closely aligned with the audio. 
The caption with the highest score is selected as the final output.
In this work, we use a CLAP model with an audio encoder HTS-AT \cite{chen2022hts} and a text encoder RoBERTa \cite{liu2019roberta}.

\section{Experimental Setup}
\subsection{Datasets}

\label{dataset}
SLAM-AAC was pre-trained using four key AAC datasets: Clotho  \cite{drossos2020clotho}, AudioCaps \cite{kim2019audiocaps}, WavCaps  \cite{mei2023wavcaps}, and MACS \cite{martin2021ground}.
For Clotho, we used version 2.1, which contains audio clips lasting 15 to 30 seconds, with captions ranging from 8 to 20 words.
This dataset includes 3,839 training, 1,045 validation, and 1,045 evaluation audio examples, each paired with five captions.
AudioCaps contains over 50,000 ten-second audio clips sourced from AudioSet \cite{gemmeke2017audio}. 
It is divided into training (49,274 clips, one caption each), validation (494 clips, five captions each), and test (957 clips, five captions each) sets.
WavCaps comprises 403,050 audio clips sourced from AudioSet, BBC Sound Effects\footnote{\scriptsize{\url{https://sound-effects.bbcrewind.co.uk}}}, FreeSound\footnote{\scriptsize{\url{https://freesound.org}}}, and SoundBible\footnote{\scriptsize{\url{https://soundbible.com}}}.
MACS includes 3,930 10-second audio files, each with 2 to 5 captions, recorded in three acoustic scenes (airport, public square, and park) from the TAU Urban Acoustic Scenes 2019 dataset.

In our experiment, the pre-training data included the training sets from Clotho, AudioCaps, and MACS, along with the entire WavCaps dataset. 
Additionally, the Clotho training set was augmented using our proposed paraphrasing method as illustrated in Section \ref{subsec:paraphrasing}.

\subsection{Evaluation Metrics}
To evaluate the quality of generated audio captions, we used several standard AAC metrics: METEOR \cite{banerjee2005meteor}, CIDEr \cite{vedantam2015cider}, SPICE \cite{anderson2016spice}, SPIDEr \cite{liu2017improved}, SPIDEr-FL \cite{zhou2022can} and FENSE \cite{zhou2022can}.
METEOR considers unigram precision, recall, synonyms, and stemming. 
CIDEr measures the consensus between generated and reference texts using TF-IDF weighted n-grams.
SPICE compares semantic graphs of generated and reference captions.
SPIDEr linearly combines CIDEr and SPICE for balanced syntactic and semantic evaluation.
SPIDEr-FL extends this by incorporating a fluency error detector from FENSE, a BERT-based model trained on audio captions, which penalizes a sentence's SPIDEr score if its fluency error probability exceeds 90\%. 
FENSE combines Sentence-BERT's semantic similarity with this fluency error detection for a comprehensive assessment.
All metrics are reported with higher values indicating better performance.

\begin{table*}[htbp]
\centering
\caption{
Ablation study of SLAM-AAC on Clotho and AudioCaps evaluation split. The ablated components are marked with \underline{underline}. PA$_{Clotho}$ refers to the paraphrasing augmentation applied to the Clotho training set during model pre-training.
} 
\resizebox{1\linewidth}{!}{
\begin{tabular}{lcccccccccccccccccccc}
\toprule
\multirow{2.5}{*}{\textbf{Ablation}} & \multicolumn{4}{c}{\textbf{Model Components}} &  & \multicolumn{6}{c}{\textbf{Clotho Evaluation} (\%)} & & \multicolumn{6}{c}{\textbf{AudioCaps Evaluation} (\%)} \\
\cmidrule(lr){2-5} \cmidrule(lr){7-12} \cmidrule(lr){14-19}
 & Audio Encoder & LLM-tuning & PA$_{Clotho}$  & Decoding & &  MT & CD & SC & SD & SF & FS && MT & CD & SC & SD & SF & FS  \\
\midrule
SLAM-AAC & EAT & LoRA & \usym{2713} & CLAP-Refine && \textbf{19.7} & \textbf{51.5} & \textbf{14.8} & \textbf{33.2} & \textbf{33.0} & \textbf{54.0} && \textbf{26.8} & \textbf{84.1} & 19.4 & \textbf{51.8} & \textbf{51.5} & \textbf{66.8}  \\
\quad w/o EAT & \underline{BEATs} & LoRA & \usym{2713} & CLAP-Refine &  & 18.5 & 47.3  & 13.7  & 30.5  &30.3 & 51.8 & & 24.6 & 74.6 & 17.7 & 46.2 & 46.0 & 63.3  \\
\quad w/o LoRA & EAT & \underline{Frozen} & \usym{2713} & CLAP-Refine &  &19.1& 47.6 & 14.4 & 31.0 &  31.0 & \textbf{54.0} && 25.7 & 76.0 & 19.1 & 47.6 & 47.3 & 66.1\\
\quad w/o PA$_{Clotho}$ & EAT & LoRA & \underline{\usym{2717}} & CLAP-Refine &  & 19.5 & 50.9& \textbf{14.8} & 32.8 & 32.7 & 53.9  && 26.6 & 81.4 & 19.4 & 50.4 & 50.2 & 66.6 \\
\quad w/o CLAP-Refine & EAT & LoRA & \usym{2713} & \underline{Beam search} && 19.4 & 51.0 & 14.6 & 32.8 & 32.7 & 53.3 && 26.1 & 82.7 & \textbf{19.5}  & 51.1 & 50.9 & 65.9  \\
\bottomrule
\end{tabular}
}
\label{tab:ablation}
\end{table*}

\subsection{Training and Inference Details}
Our model was initially trained on the pre-training datasets detailed in Section \ref{dataset} with a batch size of 16 and a peak learning rate of 1e-4, over 100,000 updates.
The learning rate was governed by a linear decay schedule, with a 1,000-iteration warmup phase before linearly decaying.
Subsequently, the model was fine-tuned separately on the Clotho and AudioCaps training sets for 10 epochs, with a reduced batch size of 4 and a peak learning rate of 8e-6.
During training, the LoRA adapter \cite{hu2021lora} was integrated into Vicuna to refine the \( q \) and \( v \)  projection layers within the Transformer \cite{vaswani2017attention} blocks. 
Model validation was conducted every 500 updates, with checkpoints saved based on the lowest validation loss.
The pre-training and fine-tuning processes were carried out on an NVIDIA A800 GPU, taking approximately 26 hours and 5 hours, respectively.

The CLAP model, used for CLAP-Refine decoding, was trained on AudioCaps, Clotho, and WavCaps, with a batch size of 128 and a peak learning rate of 5e-5 for 15 epochs. 
A cosine annealing schedule with a 2-epoch warm-up phase was employed, with the model saved at the point of lowest validation loss.

During inference, SLAM-AAC employed beam search with beam sizes ranging from 2 to 8, generating decoding candidates within each beam size based on the highest probabilities.
The final output caption is then determined using CLAP-Refine, which selects the candidate with the highest similarity score to the input audio.

\section{Experimental Results}

\subsection{Main Results}
We compared the SLAM-AAC model's performance on Clotho and AudioCaps with existing top AAC models.
Both EnCLAP \cite{kim2024enclap} and WavCaps \cite{mei2023wavcaps} utilize the supervised model HTS-AT \cite{chen2022hts} as the audio encoder and BART \cite{lewis2019bart} as the text decoder, with WavCaps benefiting from pre-training on a large-scale, weakly-labeled dataset.
Wu et al. \cite{wu2023beats} and Tang et al. \cite{tang2024extending} employ BEATs for audio feature extraction, with Tang et al. further using the large language model Vicuna \cite{vicuna2023} instead of BART for text decoding. 
Wu et al. integrate ChatGPT mixup augmentation and fine-tune their model on Clotho to validate its performance, while Tang et al. incorporate the Whisper \cite{radford2023robust} speech encoder and additional speech data \cite{panayotov2015librispeech,chen2021gigaspeech} for pre-training, validating their model on both AAC and ASR tasks.

As illustrated in Table \ref{tab:main}, SLAM-AAC achieves state-of-the-art (SOTA) performance across all metrics on both datasets.
On the Clotho dataset, SLAM-AAC slightly surpasses the previous SOTA model by Wu et al. 
On AudioCaps, it demonstrates substantial improvement ($\ge$1\%) over existing AAC systems in all metrics, indicating a closer alignment between the generated captions and human annotations in terms of semantic and lexical similarity.

\subsection{Ablation Study}
We conducted comprehensive ablation studies to validate the effectiveness of each component within SLAM-AAC.

\noindent\textbf{Network Architecture.} 
We examined the impact of different audio encoders and the necessity of fine-tuning the LLM. 
Specifically, we replaced the commonly used audio encoder BEATs with EAT, which has demonstrated superior performance in audio tagging tasks. 
This substitution resulted in a notable model performance boost, particularly on the AudioCaps evaluation, as shown in Table \ref{tab:ablation}. 
These findings suggest that the choice of audio encoder remains a critical bottleneck in AAC systems, with a more robust encoder facilitating the extraction of fine-grained audio features and enhancing the model's overall comprehension of audio.
Additionally, our ablation study on LoRA underscores the importance of efficient LLM fine-tuning. 
The incorporation of LoRA adapters proved essential for aligning the textual and audio modalities, effectively harnessing the pre-trained knowledge of the LLM for AAC tasks.

\noindent\textbf{Paraphrasing Augmentation.}
Table \ref{tab:ablation} presents the results of our ablation study on paraphrasing augmentation. 
Integrating back-translation-based text augmentation into the Clotho training dataset during pre-training notably enhanced the model's performance on both the Clotho and AudioCaps benchmarks. 
Surprisingly, the use of augmented Clotho data yielded a marked improvement in AudioCaps evaluation scores, with increases exceeding 1\% in the CIDEr, SPIDEr, and SPIDEr-FL metrics. 
These findings suggest that paraphrasing augmentation effectively enriches the data vocabulary and diversity, leading to increased model robustness and generalizability, especially when trained with limited audio-text paired data.

\begin{table}[htbp]
\centering
\caption{
Comparison of decoding strategies on AudioCaps. 
The \textcolor{gray}{gray} row denotes oracle results (theoretically optimal), derived from multiple beam search. 
Our result, which achieved the highest reranking score within CLAP-Refine, is highlighted in \textbf{bold}.
}
\label{tab:decoding_ablation}
\resizebox{1\linewidth}{!}{
\begin{tabular}{lccccccc}
\toprule
\multirow{2.5}{*}{\textbf{Decoding Strategy}} & \multirow{2.5}{*}{\textbf{Score Reranking}} & \multicolumn{6}{c}{\textbf{AudioCaps Evaluation} (\%)} \\ 
\cmidrule(lr){3-8}
& & MT & CD & SC & SD & SF & FS \\ 
\midrule
Nucleus Sampling  & - & 25.9 & 79.7 & 19.5 & 49.6 & 49.3 & 65.4 \\ 
Beam Search & - & 26.1 & 82.7 & 19.5 & 51.1 & 50.9 & 65.9 \\ 
\midrule
\multirow{4}{*}{CLAP-Refine} & \textbf{Rank 1st Set (ours)} & \textbf{26.8} & \textbf{84.1} & \textbf{19.4} & \textbf{51.8} & \textbf{51.5} & \textbf{66.8} \\
 & Rank 3rd Set & 26.2 & 81.9 & 19.3& 50.6 & 50.4 & 66.1 \\
 & Rank 5th Set & 25.7 & 80.4 & 19.2 & 49.9 & 49.6 & 65.8 \\
 & Rank 7th Set & 25.3 & 79.5 & 18.9 & 49.2 & 48.9 & 65.1 \\ 
\midrule
 \textcolor{gray}{Oracle}  & \textcolor{gray}{-} & \textcolor{gray}{27.7} & \textcolor{gray}{91.2} & \textcolor{gray}{20.3} & \textcolor{gray}{55.8} & \textcolor{gray}{55.6} & \textcolor{gray}{68.7} \\
\bottomrule
\end{tabular}
}
\end{table}

\noindent\textbf{Text Decoding Strategies.} 
Table \ref{tab:decoding_ablation} presents the comparison of different text decoding strategies.  
Results show that nucleus sampling (temperature 0.5, top-p 0.95) output yields low CIDEr and SPIDEr scores, indicating the lack of consistency with human annotations and making this approach less ideal for AAC tasks.
For beam search, captions were generated with beam sizes ranging from 2 to 8, denoted as $B_2$, $B_3$, ..., $B_8$. 
The best performance was observed with a beam size of 4 ($B_4$), which served as the benchmark for comparison.
The CLAP-Refine strategy assigns an audio-text similarity score to each beam result, reranking them based on these scores to form seven ranked sets finally, with the top-ranked set representing the final output caption.
For clarity, we display the results from the first, third, fifth, and seventh-ranked sets.
As shown in Table \ref{tab:decoding_ablation}, the higher-ranked sets consistently outperform the others, demonstrating the effectiveness of CLAP-Refine in refining beam search outputs. 
However, when compared to the theoretical optimal, represented by decoding texts with the highest FENSE score across beam results (denoted as the oracle), a substantial gap remains. 
This suggests that the optimal decoding result may span across different beam searches, indicating considerable potential for enhancing post-processing and selection strategies based on the current AAC model.

\section{Conclusion and Future Work}

In this work, we propose SLAM-AAC to enhance AAC with paraphrasing augmentation and CLAP-Refine through LLMs.
We use the EAT encoder to extract audio representations, which are then downsampled and aligned with text embeddings via linear projection layers. 
Decoding is carried out by the LLM Vicuna, with fine-tuning restricted to the projector and LoRA adapter for training efficiency. 
Drawing from back-translation in machine translation, SLAM-AAC applies paraphrasing augmentation to increase caption diversity for audio clips in Clotho, thereby improving the model's generalizability. 
Additionally, we propose CLAP-Refine, a plug-and-play decoding strategy that enhances caption selection in post-processing. 
This strategy utilizes multiple beam search outputs as candidates and selects the final caption based on the highest similarity score towards the input audio, calculated by the CLAP model.
Experiments show that SLAM-AAC outperforms existing AAC models on the AudioCaps and Clotho datasets, with ablation studies validating the contribution of each component in enhancing overall model performance.

Future work will explore alternative paraphrasing augmentation methods, such as LLM-based rephrasing, and investigate the impact of different CLAP-based models on the CLAP-Refine strategy, aiming to approach or even achieve oracle-level decoding performance.


\bibliographystyle{IEEEtran} 
\bibliography{refs}

\end{document}